\documentclass[twocolumn,showpacs,preprintnumbers,amsmath,amssymb, prb]{revtex4}
\usepackage{graphicx}
\usepackage{dcolumn}
\usepackage{bm}
\usepackage{wasysym}
\usepackage{textcomp}
\usepackage{epstopdf}
\usepackage{stmaryrd}
\usepackage{txfonts}
\begin{document}

\preprint{APS/123-QED}

\title{Self-magnetic compensation and Exchange Bias in ferromagnetic Samarium systems}
\author{P. D. Kulkarni$^1$, S. K. Dhar$^1$, A. Provino$^2$, P. Manfrinetti$^2$ and A. K. Grover$^1$}
\affiliation{$^1$Department of Condensed Matter Physics and Materials Science, Tata Institute of Fundamental Research, Homi Bhabha Road,
Colaba, Mumbai 400005, India}
\affiliation{$^2$CNR-SPIN and Dipartimento di Chimica e Chimica Industriale, Universita degli Studi di Genova, Via Dedecaneso 31, Genova 16146, Italy.}
\date{\today}

\begin{abstract}
For Sm$^{3+}$ ions in a vast majority of metallic systems, the following interesting scenario has been conjured up for long, namely, a magnetic lattice of tiny self (spin-orbital) compensated 4$f$-moments exchange coupled (and phase reversed) to the polarization in the conduction band. We report here the identification of a self-compensation behavior in a variety of ferromagnetic Sm intermetallics via the fingerprint of a shift in the magnetic hysteresis ($M$-$H$) loop from the origin. Such an attribute, designated as exchange bias in the context of ferromagnetic/antiferromagnetic multilayers, accords these compounds a potential for niche applications in spintronics. We also present results on magnetic compensation behavior on small Gd doping (2.5 atomic~\textdiscount) in one of the Sm ferromagnets (viz. SmCu$_4$Pd). The doped system responds like a pseudo-ferrimagnet and it displays a characteristic left-shifted linear $M$-$H$ plot for an antiferromagnet. 

\end{abstract}



\maketitle
\section{Introduction}

Samarium is the fifth member of the 4$f$-rare earth ($R$) series, in which spin-orbit ($S$-$L$) coupling prevails and the total angular momentum ($J$) governs the magnetic moment. In particular, for free Sm$^{3+}$ ion ($S$ = 5/2, $L$ = 5, $J$ = 5/2 and Lande's $g$-factor, $g_J$ = 2/7), the 4$f$-orbital contribution (-$\mu_B$ $\textless$$L_z$$\textgreater$ = -$\mu_B$$J$($g_J-2$)) to the magnetic moment is only 20 percent larger than its 4$f$-spin (\textbar$\mu_B$ 2$\textless$$S_z$$\textgreater$\textbar = 2$\mu_B$$J$($1-g_J$)) contribution, this renders it a small magnetic moment ($\mu_J$ = $g_J$$J$) of 5/7 $\mu_B$ \cite{McEwen}. In a crystalline environment, the degeneracy of the ground-$J$ multiplet is partially lifted, and in the specific case of the Sm$^{3+}$ ions, the crystalline electric field and the exchange field can induce admixture of the higher $J$-multiplets ($J$ = 7/2, 9/2) into the ground $J$-multiplet ($J$ = 5/2) \cite{White, Vleck, Malik, Wijn}. Such circumstances not only substantially reduce the magnetic moment/Sm$^{3+}$ ion \cite{Buschow,Malik1,Adachi1,Stewart}, but, also make the orbital and spin contributions undergo different thermal evolutions \cite{Malik, Wijn, Adachi1, Adachi2, Adachi3}. 

Enigmatic magnetic characteristics of Sm$^{3+}$ ion were first recognized by J. A. White and J. H. Van Vleck in early 1960s. \cite{White}, as they explored a ramification of the relatively small separation between the ground $J$ multiplet ($J$ = 5/2) and the first excited state ($J$ = 7/2). On allowing for the admixture of higher $J$ multiplet into the ground $J$ multiplet by the perturbation of an exchange field, they could demonstrate \cite{White} the breakdown of the proportionality between the thermal evolution of bulk magnetic susceptibility (\textless$\mu_z$\textgreater = -$\mu_R$~\textless$J_z$\textgreater) and the local 4$f$-spin susceptibility ($\alpha$\textless$S_z$\textgreater) as measured by the NMR Knight shifts. White and Van Vleck's work was extended further in 1970s (see, e.g., our Refs. [4-7] by adding the role of crystalline electric fields to the admixture effects, and it was calculated \cite{Buschow, Malik1} that the saturation magnetic moment at Sm$^{3+}$ ion could indeed vanish, with opposing orbital and spin contributions cancelling out each other, under some parametric conditions involving the crystalline electric fields and the exchange field. There were no very clear experimental elucidations of the said near cancellation.  

Our rationale for the assertion of the self-compensation behavior in Sm systems via the identification of shift in the $M$-$H$ loop from origin now stems from a recent exploration by some of us \cite{Kulkarni} in a single crystal of admixed rare earth intermetallic, viz., Nd$_{0.75}$~Ho$_{0.25}$Al$_2$, derived from a ferromagnetic $R$Al$_2$ series of compounds. It is well documented that the 4$f$-spins of the dissimilar $R^{3+}$ ions in such admixed rare earth intermetallics continue to remain ferromagnetically coupled \cite{Williams, Taylor, Grover}, which in turn mandates an anti-ferromagnetic coupling between the magnetic moments of dissimilar $R^{3+}$ ions belonging to the first half and the second half of the 4$f$-series, for which $J$ = $L$ - $S$ and $J$ = $L$ + $S$, respectively in the ground state. The Nd$^{3+}$ and Ho$^{3+}$ ions in the case cited above belong to the first/second half of 4$f$-series. The admixed Nd$_{0.75}$Ho$_{0.25}$Al$_2$ alloy, where $\mu_{Nd}$ $\approx$ 2.5 $\mu_B$ and $\mu_{Ho}$ $\approx$ 8.5 $\mu_B$, therefore responds like a ferrimagnet \cite{Swift}, and displays a \textquoteleft magnetic compensation characteristic\textquoteright. In close proximity of the compensation temperature ($T$$_{comp}$), the opposing contributions to the magnetization signal from the two pseudo-sublattices of Nd$^{3+}$ and Ho$^{3+}$ ions nearly balance out, as in a (quasi) anti-ferromagnet \cite{Kulkarni}. A very striking observation, however, is the surfacing up \cite{Kulkarni} of the left/right shift (i.e., finite exchange bias (EB) effect usually seen in ferro/antiferro-magnetic multilayers \cite{Nougues} over a small temperature interval encompassing the T$_{comp}$ in Nd$_{0.75}$Ho$_{0.25}$Al$_2$. The presence of such an EB effect in very close proximity of its $T_{comp}$, and, importantly, its sign change across it, correlates with the reversal in the orientation of the conduction electron polarization with respect to the externally applied field \cite{Kulkarni}. On drawing an analogy between the EB effect observed over a narrow interval in Nd$_{0.75}$Ho$_{0.25}$Al$_2$ with the relevant observation to be reported ahead in several Sm ferromagnets, the notion of a near self-compensation for the local moment of Sm$^{3+}$ in them would straight away become evident. We will first present the experimental results in a variety of Sm ferromagnets to validate the nomenclature of \textquoteleft near self-magnetic compensation\textquoteright. Thereafter, we shall also elucidate the phenomenon of phase reversal in the EB in one of the chosen Sm-systems, when it is substitutionally doped with few percent of Gd$^{3+}$ ions. The doped system imbibes magnetic compensation phenomenon as in pseudo-ferromagnets, which is in addition to the near self-compensation happening at Sm$^{3+}$ ion. The quasi-linear $M$-$H$ data in the doped system displays left-shift as well as right-shift from the origin as a function of temperature. 

\section{Experimental}
\subsection{Sample preparation and characterization}

Polycrystalline samples of SmCu$_4$Pd, SmPtZn, SmCd, SmZn, SmScGe and SmAl$_2$ were prepared by melting together stoichiometric amounts of the constituent elements in a sealed Tantalum crucible in an induction furnace, following usual special care and procedure for handling volatile Samarium metal \cite{Shah, Singh}. Samarium, Scandium and Gadolinium metals were of 99.9 wt.~\textdiscount\ purity. Aluminium, Copper, Zinc, Cadmium and Germanium were of 99.999 wts.~\textdiscount\ purity and Palladium and Platinum of purity 99.9 wt.~\textdiscount\ were in powder form (\textless\ 60 $\muup$m powder) to start with. While weighing, small pieces of distilled Sm and turnings of Scandium and Gadolinium were handled under pure Argon, to minimize the possibility of exposure to air. The melting in the induction furnace was slowly progressed, the melt was shaken to ensure homogenisation of the constituents. The melting stage was repeated atleast twice. To avoid the reaction of the melt to the crucible in the case of SmScGe, the binary alloy ScGe was initially prepared by arc melting and to it an appropriate amount of Sm was added, before sealing the reactants in a tantalum crucible. Different specimen were given different heating and annealing treatments, following earlier leads in the literature \cite{Adachi2, Shah, Singh}. the binary SmZn and SmCd were induction heated upto 1100 \textcelsius, followed by annealing in evacuated sealed quartz tube at 920 \textcelsius\ for 3 days. SmScGe was heated upto 1700 \textcelsius, followed by vacuum annealing at 900 \textcelsius\ for 4 days. The newly synthesized SmPtZn for the present study was induction heated upto 1400-1500 \textcelsius, followed by annealing at 800 \textcelsius\ for seven days. The samples of the admixed Sm$_{1-x}$Gd$_x$Cu$_4$Pd ($x$ = 0.01, 0.015, 0.02, 0.025, 0.03 and 0.04) were prepared by melting together appropriate amounts of the end members, SmCu$_4$Pd and GdCu$_4$Pd. All the polycrystalline samples were checked by both metallographic examination of the small pieces of the ingots and by recording x-ray diffraction patterns of all the specimen finely crushed in an inert atmosphere. As a result of the preparation method and the annealing sequence adopted, each sample was confirmed to have been formed in the desired crystallographic phase. No evidence of strong sample texturing was noted in any of the samples. In few alloy buttons, very low amount (1 to 3~\textdiscount) of extra phase(s) could, however, be seen as narrow grain separation. 

The intermetallics, SmCd and SmZn, crystallize in cubic (B2) CsCl structure \cite{Adachi1}, SmAl$_2$ and SmCu$_4$Pd have cubic MgCu$_2$ \cite{Williams}($C$15 Laves phase) and cubic MgCu$_4$Sn structures \cite{Shah}, respectively. SmScGe and SmPtZn have tetragonal CeScSi-type \cite{Singh} and  orthorhombic TiNiSi-type \cite{Dhar} structures, respectively. The intermetallic SmPtZn has been newly synthesized for the present study, its lattice parameters are $a$ = 7.015(2) $\AA$, $b$ = 4.220(1) $\AA$ and $c$ = 8.111(3) $\AA$. In SmCd and SmZn, the Sm$^{3+}$ ions form a simple cubic lattice, in SmCu$_4$Pd and SmAl$_2$, they form face centred cubic (FCC) and diamond structures, respectively. In SmScGe and SmPtZn, Sm$^{3+}$ ions yield tetragonal and orthorhombic sublattices, respectively.  
\subsection{Magnetization measurements}
For the magnetization measurements, a sample piece (20-100 mg) was cut from a given ingot and loaded in the straw holder. Magnetization measurements were performed using a SQUID-Vibrating Sample Magnetometer (Model S-VSM) of Quantum Design, Inc., U.S.A. In S-VSM, a sample vibrates along the axis of the cylindrical second derivative coil array. The SQUID sensor in S-VSM can detect a signal as low as 10$^{-7}$ emu, and the magnetization signal from the blank straw holder is $\sim$ 2 \texttimes 10$^{-7}$ emu. In the paramagnetic state, the signal from a sample of $\sim$ 20 mg is of the order of 10$^{-5}$ emu at $H$ $\sim$ 10 Oe, and this signal enhances by two-three orders of magnitude in the magnetically odered state below $T_c$. The remanent field(s) trapped inside the superconducting magnet was ascertained by recording the $M$-$H$ data while ramping the field down across the zero position from both positive and negative sides for the standard paramagnetic Pd sample. These corrections were incorporated while plotting the $M$-$H$ data for the experimental samples. We anticipate that the error in the location of the zero field position during different cycles in a given specimen is limited to within 1-2 Oe. The determined values of the effective coercive field and the exchange bias field are reliable to within 2 Oe. 

\section{Results and Discussion}

\subsection{Magnetization measurements in pristine ferromagnetic Samarium intermetallics and the fingerprint of Exchange bias}

Figure 1 shows the $M$-$H$ loops in polycrystalline samples of six Sm based ferromagnets at the temperatures as indicated. All the $M$-$H$ loops in Figs. 1(a) to 1(f) were obtained by ramping the field between \textpm~$H_{max}$, after initially cooling a given sample to a selected temperature in a field of chosen $H_{max}$. The panels (a) to (f) in Fig. 2 display the temperature variations of the magnetization ($M$ vs. $T$) measured in a field of 10 kOe in the six samples. The ferromagnetic $T_c$ values ascertained from the very low field ($H$ $\AC$ 10 Oe) $M$-$T$ curves (not shown here) in the respective samples have been indicated in the plots displayed in Fig. 2. 

The $M$-$H$ loops at $T$ $\ll$ $T_c$ in the Figs.1(a) to 1(e) are clearly asymmetric, their centres of gravity are left-shifted as the field value of crossover of the $M$ = 0 axis while ramping the field down (i.e., $H_-$) is larger in magnitude than the corresponding value while ramping the field up (i.e., $H_+$). We identify the EB field as $H_{EB}$ = -($H_-$ + $H_+$) / 2 \cite{Nougues}, so that it is deemed as positive/negative for the left/right shifted $M$-$H$ loop. The $M$-$H$ loop in Fig. 1(f) for SmAl$_2$ ($T_c$ = 125 K) at 20 K appears symmetric, however, it also yields $H_{EB}$ = +50 Oe, a left shifted value, much lower than the $H_{EB}$ values indicated in the Figs. 1(a) to 1(e) in the other Sm-ferromagnets. $H_{EB}$ value at a given temperature in a given compound depends somewhat on the value of initial cooling field and the chosen $H_{max}$. Typically, upto 10 \textdiscount\ change in a given $H_{EB}$ value could be noted on changing the initial field in which a sample is cooled and thereafter varying the $H_{max}$ upto 70 kOe, the highest field available to us in the Quantum Design Inc. S-VSM system.

\begin{figure}[h]
\includegraphics[width=0.45\textwidth]{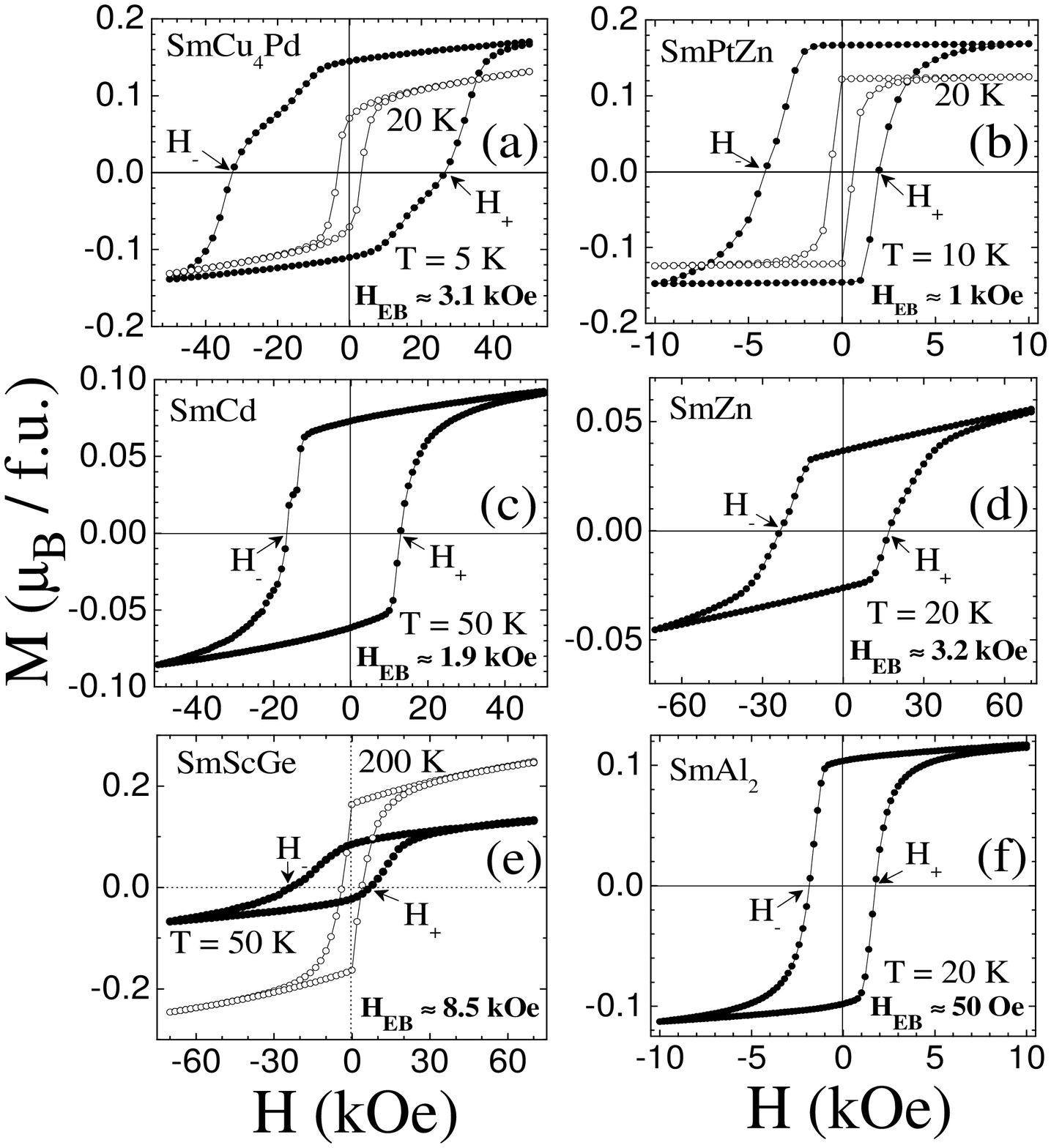}
\caption{\label{fig1} $M$-$H$ loops at selected temperatures (\textless~$T_c$) in, (a) SmCu$_4$Pd, (b) SmPtZn, (c) SmCd, (d) SmZn, (e) SmScGe and (f) SmAl$_2$. The exchange bias field ($H_{EB}$ = -($H_-$ + $H_+$)/2), which measures the left shift in a given $M$-$H$ loop at a given temperature, is stated in each of the panels, a to f.}
\end{figure}

\begin{figure}[h]
\includegraphics[width=0.45\textwidth]{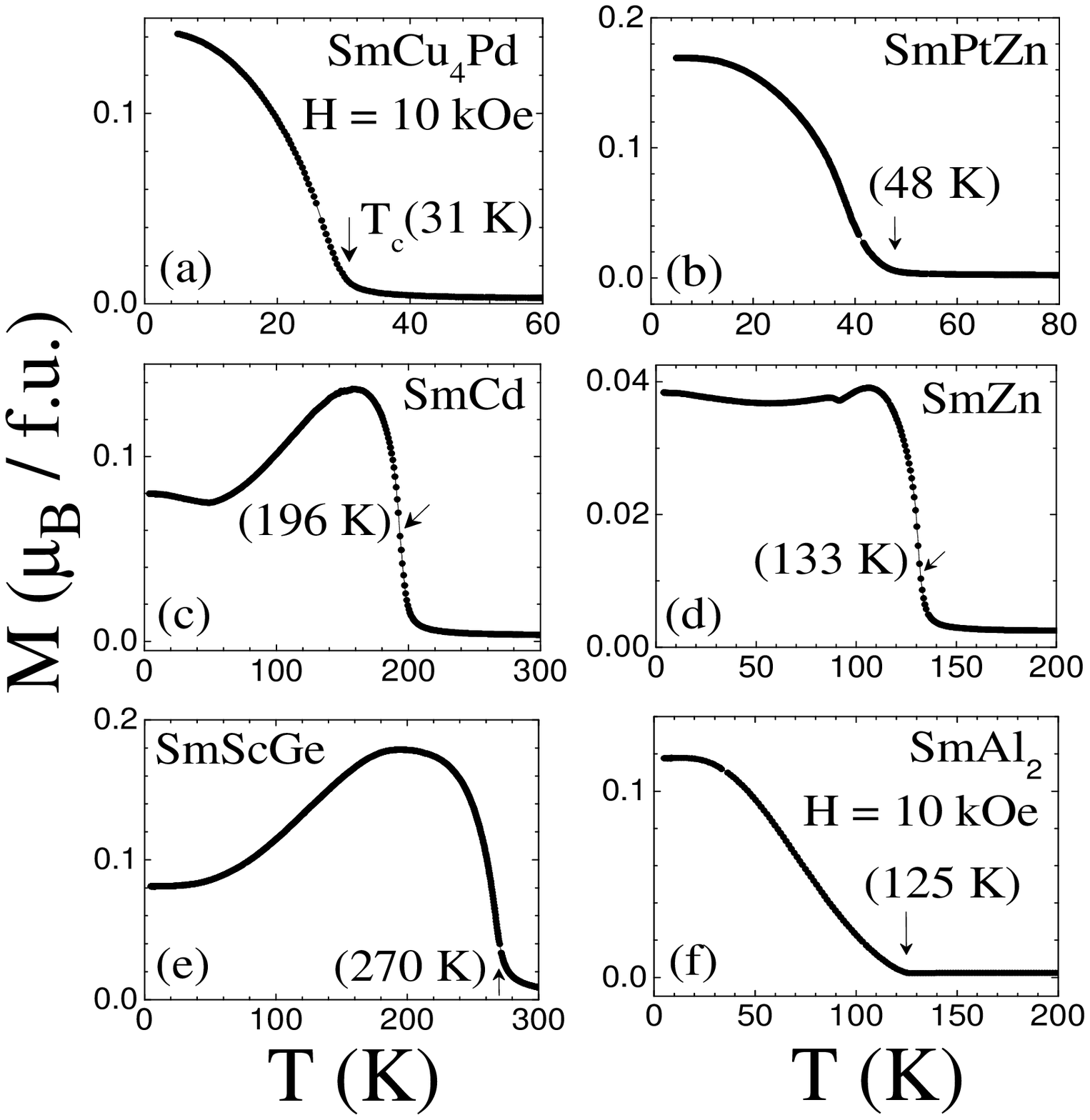}
\caption{\label{fig2} $M$ vs $T$ measured while cooling down in a field of 10 kOe in different Sm compounds. $T_c$ values determined from the low field ($H$ $\AC$ 10 Oe) $M$-$T$ curves (not shown here) are indicated with arrows.}
\end{figure}

It is fruitful to examine the small magnetic moment values in Figs. 1(a) to 1(f), in addition to the presence of EB effect in them. Adachi $et~al.$ \cite{Adachi1} had determined conduction electron polarization (CEP) in SmCd, SmZn and SmAl$_2$ to be antiferromagnetically linked to the net local moment of the Sm$^{3+}$ ions. The CEP contributions at $T$ = 0 K range between 0.2 and 0.4 $\mu_B$, whereas Sm magnetic moment values are calculated \cite{Adachi1} to lie between 0.3 to 0.5 $\mu_B$. In ferromagnetic GdCu$_4$Pd and GdScGe, the magnetic moment per formula unit values are above 7.6 $\mu_B$ \cite{Shah, Ivanova}, thereby, elucidating that the contribution from polarized conduction electron adds on to the local moment of 7 $\mu_B$/Gd$^{3+}$ in them. This in turn implies that the CEP contribution in SmScGe and SmCu$_4$Pd would also be in phase \cite{Stewart} with the contribution from the 4$f$-spin of Sm$^{3+}$, and out of phase with net 4$f$-magnetic moment of Sm$^{3+} ion$. Continuing in the same vein, note further the nearly saturated value of 0.2 $\mu_B$/formula unit ($f.u.$) in newly synthesized SmPtZn in Fig. 1(b); its smallness once again attests to the competition between the contributions to the magnetization signal from the local 4$f$-magnetic moment of Sm$^{3+}$ and that from the CEP.

\begin{figure}[h]
\includegraphics[width=0.3\textwidth]{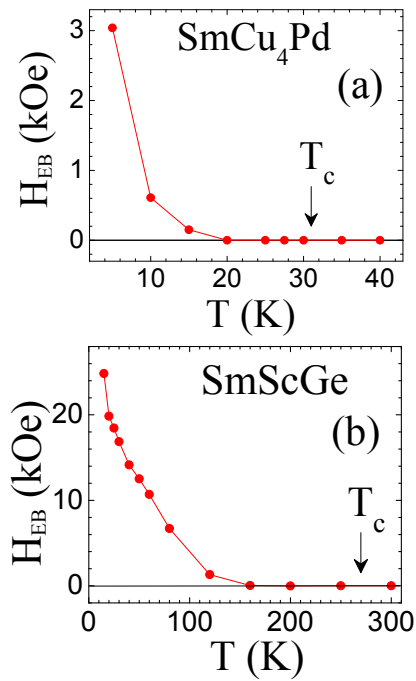}
\caption{\label{fig3} (Color online) Temperature variation of left shift, i.e., exchange bias field in ferromagnets, (a) SmCu$_4$Pd and (b) SmScGe. The respective $T_c$ values are indicated with arrows.}
\end{figure}

Another instructive observation is the contrast in the shape of $M$-$T$ curves in Figs. 2(a), 2(b) and 2(f) with that of the $M$-$T$ curves in Figs. 2(c), 2(d) and 2(e). The former shape reflects the monotonic increase in the magnetization expected in ferromagnets, whereas the latter non-monotonic peak like response is akin to that usually seen in ferrimagnets. We recall here that Adachi $et~al.$ \cite{Adachi1} pointed out that while SmAl$_2$ (cf. Fig. 1(f)) is a usual \textquoteleft orbital surplus\textquoteright\ Sm ferromagnet; in SmCd (cf. Fig. 1(c)) as well as in SmZn (cf. Fig. 1(d)), the contribution to the total magnetization from the local magnetic moment of Sm$^{3+}$ at $T$ = 0 K was calculated to be only ($\AC$ 0.35 $\mu_B$), which is smaller than the calculated contribution ($\AC$ 0.4 $\mu_B$) from the CEP. The local moment contribution is phase reversed to the CEP contribution in Sm-ferromagnets under study. The latter two compounds, viz., SmCd and SmZn, were thereby termed as \textquoteleft spin-surplus\textquoteright\ Sm-ferromagnets \cite{Adachi1}. In a \textquoteleft spin-surplus\textquoteright\ Sm ferromagnet, the sum of the contributions from the local 4$f$-spin moment and the CEP exceeds that from the local 4$f$-orbital contribution. The opposing \textquoteleft spin\textquoteright\ and \textquoteleft orbital\textquoteright\ contributions thermally evolve to their respective saturated values at $T$ = 0 in a monotonic manner on cooling below the $T_c$, however, the former does so faster than the latter. Such a difference in the temperature dependences of the \textquoteleft spin\textquoteright\ and the \textquoteleft orbital\textquoteright~contributions can yield a peak like characteristic in the $M$-$T$ curve of a \textquoteleft spin-surplus\textquoteright\ Sm ferromagnet \cite{Adachi1, Adachi2}. It is worth reiterating that CEP contribution is aligned along the applied field in the \textquoteleft spin-surplus\textquoteright\ situation and phase reversed to it in the \textquoteleft orbital-surplus\textquoteright\ ferromagnets. Extending the reasoning further, the differences in the contours of the $M$-$T$ curves in Figs. 2(a), 2(b) and 2(f) in the \textquoteleft orbital-surplus\textquoteright~cases also reflect the differences in the thermal evolution of contributions from the 4$f$-orbital and the 4$f$-spin parts to the magnetization of the Sm$^{3+}$ ion in different crystallographic environment and under the influence of different exchange field values.  

Figures 3(a) and 3(b) summarize the temperature variations of exchange bias field ($H_{EB}$($T$) in the \textquoteleft orbital-surplus\textquoteright\ SmCu$_4$Pd ($T_c$ $\approx$ 31 K) and \textquoteleft spin-surplus\textquoteright\ SmScGe ($T_c$ $\approx$ 270 K), systems. It can be immediately seen that the exchange bias field values start to build up well below the respective $T_c$ values. Our premise that an exchange bias fingerprints the underlying approach to compensation between the local 4$f$-orbital and 4$f$-spin contribution to the magnetic moment of Sm$^{3+}$, therefore, implies that it takes a while before the near self-compensation stage gets attained on cooling below $T_c$. Note further that the $H_{EB}$ values in both the compounds remain positive down to the lowest temperature. In SmCu$_4$Pd, $H_{EB}$ values build upto 3 kOe at 5 K, whereas, in SmScGe, these values build upto 25 kOe at 15 K. 

\subsection{Magnetic compensation phenomenon in a doped Samarium ferromagnet and the Exchange Bias behavior}

It is tempting to conceive a pristine Sm ferromagnet which can display in its thermomagnetic response at low fields the crossover of $M$ = 0 axis at $T_{comp}$ [see e.g., the calculated $M$-$T$ curves in Fig. 1 of Ref. 9], and also exhibit a change in sign of $H_{EB}$ to the negative values across its $T_{comp}$ (similar to the one reported in Nd$_{0.75}$Ho$_{0.25}$Al$_2$ \cite{Kulkarni}). However, to the best of our knowledge magnetic compensation phenomenon has so far not been reported in any pristine Sm-ferromagnet. In this context, it is fruitful to explore the occurrence of magnetic compensation behavior and the crossover of $H_{EB}$ to the negative values in SmCu$_4$Pd compound by substituting few percent of the Sm$^{3+}$ by the $S$-state ($L$ = 0) Gd$^{3+}$ ions. In the doped samarium alloys, the ferromagnetism between the 4$f$-spins of Sm$^{3+}$ and Gd$^{3+}$ ions prevails~\cite{Grover}. 

\begin{figure}[h]
\includegraphics[width=0.40\textwidth]{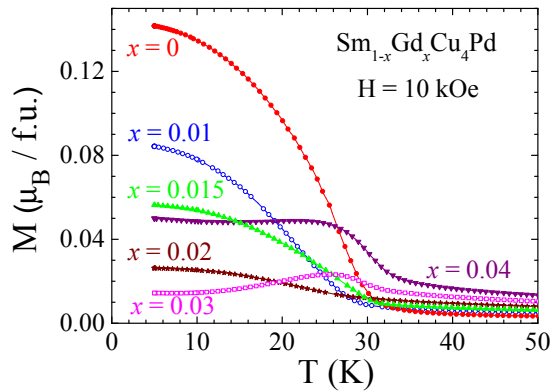}
\caption{\label{fig4} (Color online) $M$ vs $T$ measured while cooling down in a field of 10 kOe in  Sm$_{1-x}$Gd$_{x}$Cu$_4$Pd series with $x$ = 0, 0.01, 0.015, 0.02 and 0.03.}
\end{figure}

The main panel of Fig. 4 shows the $M$($T$) curves at $H$ = 10 kOe in spin-ferromagnetic Sm$_{1-x}$Gd$_x$Cu$_4$Pd alloys for 0 $\leq$ $x$ $\leq$ 0.04. Note that the magnetization value per $f.u.$ at the lowest temperature decreases from $x$ = 0 to $x$ = 0.03. The substitution of Sm$^{3+}$ by Gd$^{3+}$ ions progressively enhances the overall 4$f$-spin contribution vis-a-vis the opposing 4$f$-orbital contribution in the Sm$_{1-x}$Gd$_x$Cu$_4$Pd matrix. At $x$ = 0.02, the $M$($T$) response from the $T_c$ end monotonically decreases, whereas the $M$($T$) curve for $x$ = 0.03 has a peak like shape, akin to that associated with a \textquoteleft spin-surplus\textquoteright~Sm-ferromagnet. These data imply that the magnetic compensation behavior can be observed between $x$ = 0.02 and $x$ = 0.03. 
Figure 5 summarizes the magnetization data in $x$ = 0.025 sample. Figure 5(a) depicts the $M$($T$) curves in $H$ = 50 Oe and 2 kOe, respectively. The crossover of the $M$ = 0 axis at $T_{comp}$ = 21 K in the lower field $M$($T$) curve, and the turnaround behavior in the higher field curve at about 21 K in Fig. 5(a) characterize the occurrence of magnetic compensation.  

\begin{figure}[h]
\includegraphics[width=0.45\textwidth]{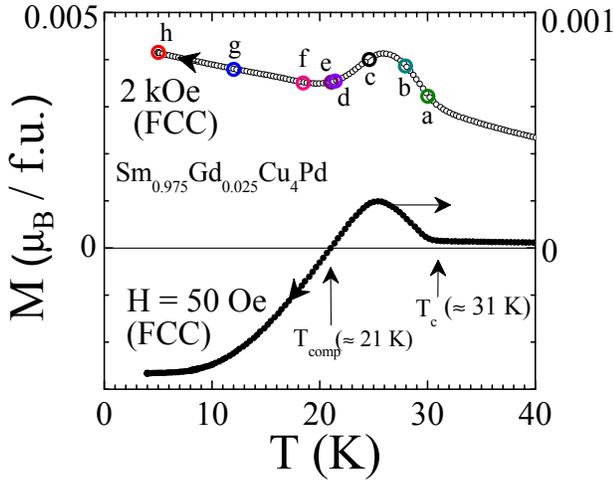}
\caption{\label{fig5} (Color online) $M$ vs $T$ in Sm$_{0.975}$Gd$_{0.025}$Cu$_4$Pd at $H$ = 50 Oe and 2 kOe. The $T_{c}$ and $T_{comp}$ values are marked with arrows in low field FCC curve. The encircled points (from (a) to (h)) on FCC curve at 2 kOe correspond to the temperatures where the $M$-$H$ loops are traced as shown later in Fig. 6.}
\end{figure}

\begin{figure}[h]
\includegraphics[width=0.45\textwidth]{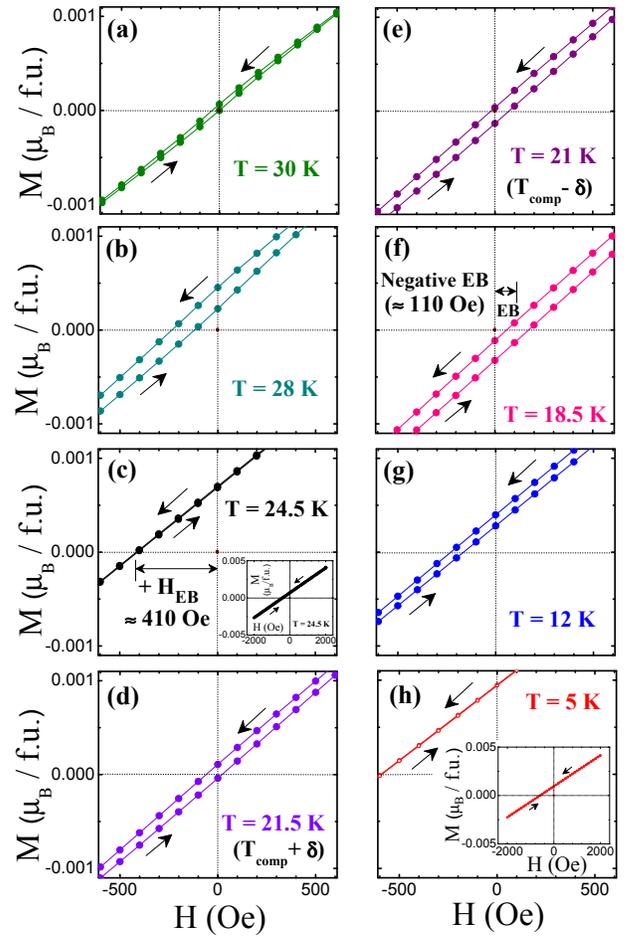}
\caption{\label{fig6} (Color online) Portions of $M$ vs $H$ loops in Sm$_{0.975}$Gd$_{0.025}$Cu$_4$Pd $M$ at eight different temperatures as indicated in panels (a) to (h). The sample is cooled from above $T_c$ ($\approx$ 31 K) down to the selected temperature and the loop is traced between \textpm~2 kOe in each case.}
\end{figure}

A set of eight panels in Fig. 6 show the portions of the $M$-$H$ loops between \textpm\ 600 Oe in Sm$_{0.975}$Gd$_{0.025}$Cu$_4$Pd sample at the temperatures identified in Fig. 5 (see the encircled positions, (a) to (h), in this panel). The sample was initially cooled down in a field of 2 kOe to each temperature, and thereafter the field was ramped between \textpm\ 2 kOe. The inset panels in Fig. 6(c) and Fig. 6(h) show the representative $M$-$H$ data between \textpm\ 2 kOe at 24.5 K and 5 K, respectively. Note first the linear nature of the $M$-$H$ response between \textpm 2 kOe over the entire temperature range in the magnetically ordered state. This feature, combined with the smallness of the magnetization values, brings out the quasi-antiferromagnetic response at this stoichiometry. This is in complete contrast to the ferromagnetic shape of the $M$-$H$ response (alongwith the large magneto-crystalline anisotropy and the associated coercivity) in the parent SmCu$_4$Pd compound (see Fig. 1(a)). However, the shift in the centre of the gravity (CG) of the $M$-$H$ data can be noted in different panels of Fig. 6. A tiny left-shift in the CG of the $M$-$H$ data surfaces up at 30 K itself (cf. Fig. 6(a)), which is just below the $T_c$ value of 31 K. The left shift progressively enhances as the temperature lowers further to 24.5 K (cf. Fig. 6(c)). Interestingly, at 24.5 K, the width of the loop has collapsed. On further lowering the temperature, left-shift in CG of the $M$-$H$ data decreases, whereas its width enhances. On lowering temperature below about 21 K ($\approx$ $T_{comp}$), the left shift crosses over to the right shift (cf. Fig. 6(d) and Fig. 6(e)). The right shift (negative EB) maximizes at 18.5 K (cf. Fig. 6(f)) and, thereafter, it starts to swing backwards. On lowering temperature below about 16 K (cf. Fig. 6(g)), the CG of the $M$-$H$ data returns to the left of the origin (cf. Fig. 6(g) at 12 K). At 5 K, $M$-$H$ data once again reaches the collapsed stage (cf. Fig. 6(h)) and $H_{EB}$ value reaches upto $\sim$570 Oe. 

The two panels ((i) and (ii)) in Fig. 7 summarize the $H_c$$^{eff}$ (the effective coercive field\cite{Webb} defined as $H_c$$^{eff}$ = -($H_-$ - $H_+$) / 2) and $H_{EB}$ values determined from the $M$-$H$ plots of the kind plotted in Fig. 6. The encircled temperatures in the two panels of Fig. 7 help to identify the respective panels in Fig. 6. It is satisfying to note the ocurrence of change in sign of $H_{EB}$ across $T_{comp}$ ($\approx$ 21 K) in panel (ii) of Fig. 7; this characteristic is consistent with the similar behavior across its $T_{comp}$ in the single crystal of Nd$_{0.75}$Ho$_{0.25}$Al$_2$. In the latter system, measureable $H_{EB}$($T$) values exist only in the neighbourhood of $T_{comp}$, where the contributions from the local moments of Nd and Ho ions nearly compensate each other. The observation that appreciable $H_{EB}$($T$) values exist over the entire temperature range below $T_c$ in Sm$_{0.975}$Gd$_{0.025}$Cu$_4$Pd implies that compensation between the orbital contribution to the magnetic moment of Sm$^{3+}$ ions and weighted average of spin contributions of local moments of Sm$^{3+}$ and Gd$^{3+}$ ions prevails throughout in the magnetically ordered state. The character of this is quasi-antiferromagnetic, as evidenced by the linear $M$-$H$ data in Fig. 6.  

\begin{figure}[h]
\includegraphics[width=0.45\textwidth]{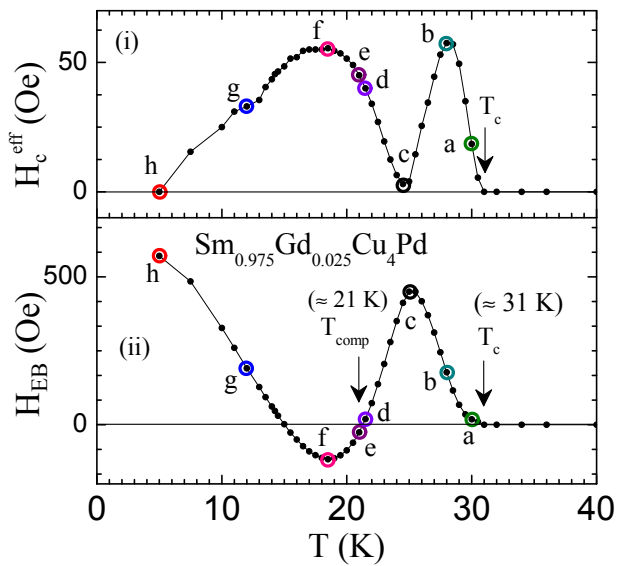}
\caption{\label{fig7} (Color online) Temperature variation of (a) $H_c$$^{eff}$ and (b) $H_{EB}$ in Sm$_{0.975}$Gd$_{0.025}$Cu$_4$Pd. The values of $H_c$$^{eff}$ and $H_{EB}$ are obtained from the $M$ vs $H$ loops of the kind shown in Fig. 6 at various temperatures below $T_c$ (marked with arrow in both panels). The encircled positions in panel (a) and (b) display the selected temperatures at which portions of $M$ vs $H$ loops have been displayed in Fig. 6.}
\end{figure}

The effective coercive field ($H_c$$^{eff}$) in Sm$_{0.975}$Gd$_{0.025}$Cu$_4$Pd in panel (i) of Fig. 7 registers the collapse in the width of the $M$-$H$ loop at 24 K, instead of at its $T_{comp}$ value, as noted in the single crystal of Nd$_{0.75}$Ho$_{0.25}$Al$_2$~\cite{Kulkarni} and polycrystalline samples of Pr$_{0.8}$Gd$_{0.2}$Al$_2$ and Sm$_{0.98}$Gd$_{0.02}$Al$_2$\cite{Kulkarni1}. It is useful to recall here that for a soft ferromagnet in contact with a ferrimagnet, Webb et al. \cite{Webb} had argued that, the effective coercivity diverging from the low as well as high temperature ends should collapse on approaching the compensation temperature of the ferrimagnet. The collapse of $H_c$$^{eff}$ at 24.5 K (\textgreater~$T_{comp}$) in the present compound is therefore curious. The location of peak in $H_{EB}$($T$) at 24.5 K is also a surprise. In any case, the left-shifted linear $M$-$H$ plots at 24.5 in Fig. 6(c) vividly elucidates the exchange bias notion for an antiferromagnetic system. The Sm moment in the pristine SmCu$_4$Pd compound is nearly self-compensated below about 20 K (cf. Fig. 3), however, the system retains its 'orbital-surplus' ferromagnetic characteristic over the entire temperature range below $T_c$. The substitution of 2.5 atomic~\textdiscount\ of Sm$^{3+}$ by Gd$^{3+}$ ions in the doped alloy makes it a pseudo-ferrimagnet. Between $T_c$ and $T_{comp}$ of 21 K, the doped alloy is prima facie \textquoteleft spin-surplus\textquoteright\. The orbital contribution of Sm$^{3+}$ ion overcomes the weighted average of spin-contributions of Sm$^{3+}$ and Gd$^{3+}$ ions alongwith that of the conduction electron polarization only below $T_{comp}$ of 21 K. The presence of second crossover in $H_{EB}$($T$) in Sm$_{0.975}$Gd$_{0.025}$Cu$_4$Pd near 15 K appears to echo the two crossovers in $H_{EB}$$(T)$ noted earlier in the compensated polycrystalline samples of Pr$_{0.8}$Gd$_{0.2}$Al$_2$ and Sm$_{0.98}$Gd$_{0.02}$Al$_2$ as well \cite{Kulkarni}. The complex $H_{EB}$$(T)$ response in Fig. 7 perhaps arises from an interplay between the approach to self-compensation happening at each Sm$^{3+}$ ion and the competition between the weighted average of residual magnetic moment associated with host Sm$^{3+}$ ions and that of the doped Gd$^{3+}$ ions in the background of polarized conduction band, whose own contribution is an intimate participant in the ongoing competition. We believe that at 24.5 K, (1-x)~\textbar$\muup$$_{Sm}$\textbar\ $\approx$ x~\textbar$\muup$$_{Gd}$\textbar\ and at $T_{comp}$ of 21 K, in low fields, (1-x)~\textbar$\muup$$_{Sm}$\textbar\ + \textbar$\muup$$_{Gd}$\textbar\ + \textbar$\muup$$_{CEP}$\textbar\ $\approx$ 0, where \textbar$\muup$$_{CEP}$\textbar\ is the contribution from the conduction electron polarization. In a larger field (e.g., in 2 kOe), the CEP contribution turns around with respect to the applied field and this could account for the change in sign of $H_{EB}$ across $T_{comp}$. The rationalization of the second crossover in $H_{EB}$($T$) at 15 K perhaps requires further exploration in a single crystal sample of the doped Sm-system.

\section{Summary and Conclusion}
To summarize, we have searched and identified exchange bias effect in the $M$-$H$ loops of several ferromagnetic Samarium intermetallics having different crystallographic structures. Small (almost) self-compensated local moments of Sm$^{3+}$ ions occupy a unique site in these ferromagnets. The net magnetization values per formula unit in these compounds are small as the conduction electron polarization competes as well with the contribution from the tiny local moment of Sm$^{3+}$ ions. One of the ferromagnetic Sm compounds is driven to yield a magnetic compensation characteristic by substitutional replacement of few percent of Sm$^{3+}$ by the Gd$^{3+}$ ions. In the compensated stoichiometry, the EB field is elucidated to change sign across $T_{comp}$ and echo the recent findings \cite{Kulkarni, Kulkarni1}. The behavior in the Sm compounds is generic, and has the potential for niche applications of magnetically ordered materials having large spin polarization, negligible self-stray field and the tunable exchange bias. A spin-valve device in spintronics typically incorporates a free ferromagnetic layer and a pinned ferromagnetic layer, exchange coupled to an antiferromagnetic bilayer \cite{Kanai}. In such a device, the pinned multi-layer portion of the composite could be considered for replacement by a layer of the zero-magnetization Sm ferromagnet.

\textbf{Acknowledgements}

We acknowledge fruitful discussions with S. Ramakrishnan, S. Venkatesh, A. Thamizhavel, Vikram Tripathi, Gianni Blatter, T. Nakamura, H. Suderow. We thank U. V. Vaidya for his help with magnetization measurements.

\end{document}